\def\year{2022}\relax
\documentclass[letterpaper]{article}
\usepackage{aaai22}  
\usepackage{times}  
\usepackage{helvet}  
\usepackage{courier}  
\usepackage[hyphens]{url}  
\usepackage{graphicx} 
\def\UrlFont{\rm}  
\usepackage{natbib}  
\usepackage{caption} 
\DeclareCaptionStyle{ruled}{labelfont=normalfont,labelsep=colon,strut=off} 
\usepackage{algorithm}
\usepackage{algorithmic}
\usepackage[switch]{lineno} 
\usepackage{newfloat}
\usepackage{listings}
\floatstyle{ruled}
\newfloat{listing}{tb}{lst}{}
\floatname{listing}{Listing}
\newcommand{\cmark}{\ding{51}}%
\newcommand{\xmark}{\ding{55}}%
\newcommand{\Larg}{\mathcal{L}}
\usepackage{times}
\usepackage{epsfig}
\usepackage{graphicx}
\usepackage{amsmath}
\usepackage{amssymb}
\usepackage{algorithm}
\usepackage{algorithmic}
\usepackage{amssymb}
\usepackage{pifont}
\usepackage{comment}
\usepackage{multirow}
\usepackage{booktabs}
\usepackage{bm}
\usepackage{graphicx}
\usepackage{multirow}
\usepackage{color}
\usepackage{subfig}

\title{Preserving Privacy in Federated Learning with Ensemble \\ Cross-Domain Knowledge Distillation }
\author {
    Xuan Gong\textsuperscript{\rm 1},
    Abhishek Sharma\textsuperscript{\rm 2},
    Srikrishna Karanam\textsuperscript{\rm 2},
    Ziyan Wu\textsuperscript{\rm 2}, \\
    Terrence Chen\textsuperscript{\rm 2},
    David Doermann\textsuperscript{\rm 1},
    Arun Innanje\textsuperscript{\rm 2}
}
\affiliations {
    \textsuperscript{\rm 1} University at Buffalo, Buffalo NY\\
    \textsuperscript{\rm 2} United Imaging Intelligence, Cambridge MA\\
    xuangong@buffalo.edu, 
    abhishek.sharma@uii-ai.com,
    srikrishna.karanam@uii-ai.com,
    ziyan.wu@uii-ai.com,
    terrence.chen@uii-ai.com,
    doermann@buffalo.edu,
    arun.innanje@uii-ai.com
}
\usepackage{bibentry}

\begin{document}
\maketitle

\begin{abstract}
Federated Learning (FL) is a machine learning paradigm where  local nodes collaboratively train a central model while the training data remains decentralized. Existing FL methods typically share model parameters or employ co-distillation to address the issue of unbalanced data distribution. However, they suffer from communication bottlenecks. More importantly, they risk privacy leakage. In this work, we develop a privacy preserving  and communication efficient method in a FL framework with one-shot offline knowledge distillation using unlabeled, cross-domain public data. We propose a quantized and noisy ensemble of local predictions from completely trained local models for stronger privacy guarantees without sacrificing accuracy. Based on extensive experiments on image classification and text classification tasks, we show that our privacy-preserving method outperforms baseline FL algorithms with superior performance in both accuracy and communication efficiency.
\end{abstract}

\section{Introduction}
The availability of large collections of data has facilitated the recent success of deep learning. However, in many cases, this wealth of data is dispersed over numerous physical locations and controlled by separate entities. Consequently,  collaboration among parties, especially clinical institutions, is restricted due to the decentralized nature of the data.  This is especially true for medical images where various legal, privacy, technical, and data ownership concerns often make it impractical or even impossible to gather such medical data to a centralized location.

To tackle some of these issues, federated learning (FL)~\cite{shokri2015privacy,yang2019federated} has emerged as a practical machine learning paradigm where local models are used to collaboratively train a centralized model using data-free communication. There are several important challenges that make FL markedly different than typical distributed learning.  First, privacy is a key concern.  It is essential that local data remain protected. Second, communication is a critical bottleneck, so steps must be taken to minimize its detrimental effects. Third, due to the decentralized nature of the collection (leading to different settings), data across various local parties are typically heterogeneous, rendering the typical machine learning assumption of independent and identical distributions (i.i.d.) invalid.

\begin{figure*}
\begin{center}  
\includegraphics[width=0.9\linewidth]{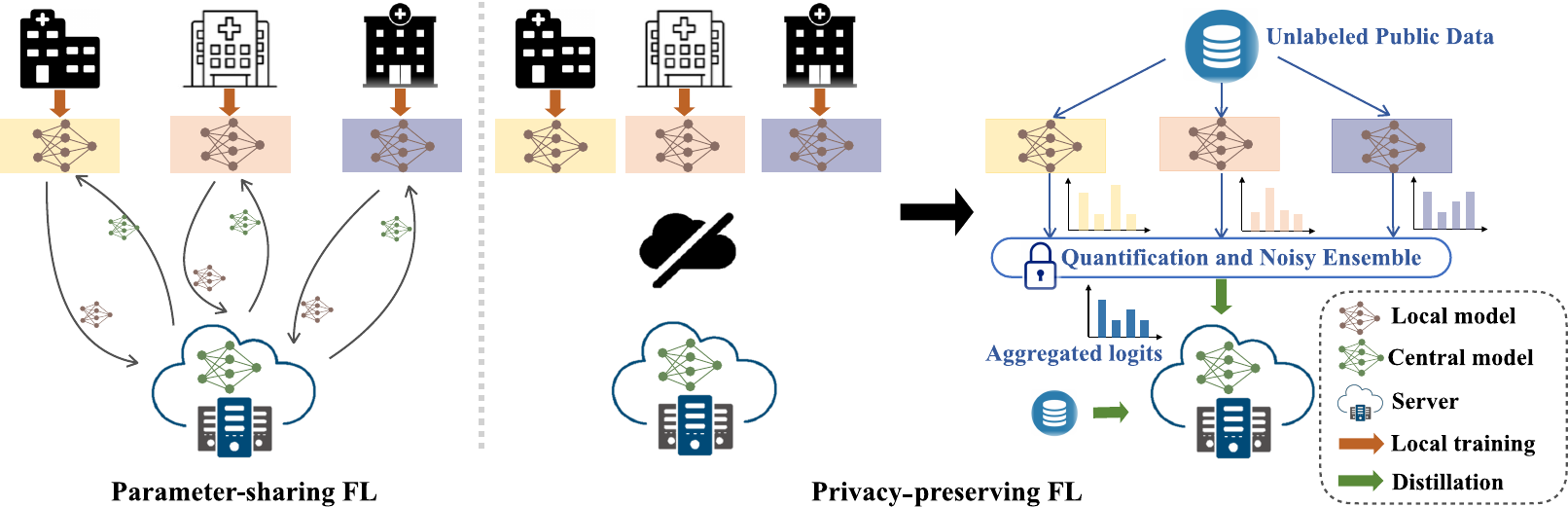}
\end{center}
   \caption{Traditional methods transfer private parameters or gradients from local nodes to a server, risking privacy leakage.  Our framework trains local models independently, and only transfers products of the unlabeled public data. We further perturb the local predictions with a quantized and noisy ensemble for a stronger privacy guarantee.}
\label{fig1}
\end{figure*}

Mainstream federated learning methods are based on the repeated sharing of parameters or gradients of local models during the training process~\cite{mcmahan2017communication,smith2017federated,li2018federated,zhao2018federated,hsu2019measuring,wang2020federated,karimireddy2019scaffold}. 
Typically, such approaches involve each local model sharing its gradients with a central server after each round of local training on its local data.  The central server then aggregates the local model parameters with typical data aggregation techniques~\cite{wang2020federated,li2019fair,hsu2020federated}. Each local node then updates its local model with the latest global aggregation, and this process continues. These parameter-based communication methods have many known security weaknesses and are limited only to models with homogeneous architectures. While some methods have shown hope of protecting against data leakage in medical imaging \cite{li2019privacy,li2020multi}, sharing parameters/gradients is highly susceptible to privacy leakage, and stealthy attacks. Some recent works ~\cite{zhu2019deep, geiping2020inverting} demonstrated the ability to obtain local private data from publicly shared gradients, further highlighting the associated privacy risks in general and in medical applications in particular. 

Another class of approaches in FL is to fuse local models into a single central model based on knowledge distillation~\cite{hinton2015distilling}.  Knowledge distillation  eliminates the requirement for identical model architectures. 
While \cite{li2019fedmd} distills the locally-computed knowledge on auxiliary public data to get around data privacy issues, they assume both the public and private data are sampled from the same underlying distribution. This further exposes the private data to security attacks. Recently proposed FedDF~\cite{lin2020ensemble} relaxes the public data to be unlabeled and non-sensitive (i.e., sampled from another domain). Similarly, \cite{zhu2021data} eliminates the prerequisite of public data with a generator and aggregates knowledge in a data-free manner. However, both of them still exchange model parameters recursively, resulting in privacy vulnerabilities due to model memorization~\cite{zhu2019deep}. 

To address these issues, we present a new framework for federated learning (Fig.~\ref{fig1}) with several innovations. First, unlike existing FL methods for either  general
or medical
~applications, our framework only shares the outputs of public data with one-shot (single round) distillation.  The public data is unlabeled and decoupled from the private data. This, by design, eliminates the security vulnerabilities identified in prior works.
Second, in contrast to the existing distillation-based FL work~\cite{li2019fedmd,lin2020ensemble, sui2020feded, zhu2021data} that exclusively train local models incrementally and update them synchronously through online distillation, we keep the local training asynchronous and independent,  and then aggregate the local predictions on unlabeled cross-domain public data.  This offline strategy largely limits the server's exposure to local models' knowledge, reducing the consumption of communication bandwidth and reducing the risk of local information leakage. Furthermore, we deploy quantized and noisy aggregation on the locally-computed logits for stronger privacy guarantees. 
We experiment with CIFAR10/100 and large-scale chest x-ray datasets, showing very competitive classification results in accuracy, bandwidth, and privacy guarantees. Extensive experiments on text classification tasks also demonstrate our method outperforms prior works with higher accuracy, lower bandwidth as well as stronger privacy guarantee. 

Our contributions can be summarized as:
\begin{itemize}
    \item We propose a one-shot federated learning framework with one-way knowledge distillation (FedKD) on unlabeled, cross-domain, non-sensitive public data, explicitly addressing the communication bottleneck and preserving the privacy of local proprietary data without sacrificing accuracy. 
    \item We introduce a seminal quantized and noisy ensemble before distillation, so that the privacy cost is meaningfully decreased with stronger security guarantees. 
    \item We demonstrated the flexibility and efficiency of the proposed framework with extensive evaluations showing superior performance on accuracy, bandwidth and privacy-preserving capability compared to prior arts,  on both image classification and text classification tasks. 
\end{itemize}

\section{Related Work}
\subsection{Knowledge Ensemble}
With the success of knowledge transfer~\cite{hinton2015distilling}, recent advancements on ensemble networks are dominated by the student-teacher learning paradigm~\cite{shazeer2017outrageously, ZhouSCZWYZ21, SongW0ZLY21}. Ensemble learning aggregates the knowledge of multiple teachers before it distills the knowledge into the student network. 
Supervised ensemble learning  is dominated by gate learning to design the weight for aggregation~\cite{shazeer2017outrageously, asif2019ensemble, xiang2020learning}. In semi-supervised and self-supervised scenarios, \cite{wu2019distilled} and \cite{you2017learning} exploit the relative similarity between samples for aggregation weights. Furthermore, co-distillation extends one-way transfer to bidirectional collaborative learning~\cite{song2018collaborative, zhu2018knowledge, dvornik2019diversity, guo2020online}. 

\subsection{Federated Learning.}
In parameter-based FL methods, each local model shares its parameters/gradients with the central server after every round of local training on its local data, following which the central server aggregates them by average~\cite{mcmahan2017communication}. The result of this aggregation step is then shared by the central server with the local nodes, which in turn update their corresponding local model and proceed with the next training round. This process is then repeated until the stopping criterion is met. A variety of extensions of FedAVG~\cite{mcmahan2017communication, wang2020federated, li2019fair, hsu2020federated} employ improved aggregation schemes, such as adding momentum~\cite{hsu2019measuring}, and local weighting~\cite{li2019fair, hsu2020federated}. Another set of approaches improve local training by incorporating proximal term~\cite{li2018federated} or control variations~\cite{karimireddy2019scaffold} to restrict local training.
However, such sharing of model parameters or gradients can be thought of as a naive way of information exchange, it is highly susceptible to privacy leakage and stealth attacks, as also demonstrated elsewhere~\cite{zhu2019deep,geiping2020inverting}.

Federated distillation methods exchange model outputs rather than model parameters. Given that some methods  produce central models by distilling knowledge from private data ~\cite{zhou2020distilled, shin2020xor} in the same spirit as those above, there is a growing concern on local data privacy. In contrast, some works~\cite{jeong2018communication, li2019fedmd} distill with the output of public data.
Although model agnostic, these methods select public data based on the prior knowledge of private data. 
Recently proposed methods FedDF~\cite{lin2020ensemble} and FedGEN~\cite{zhu2021data} relax the prerequisites of distillation data, but they are still far from privacy-preserving or communication efficient due to the iterative exchange of models over hundreds of rounds. Besides, the above mentioned approaches exclusively require many rounds of back-and-forth communication, leading to bandwidth bottlenecks and other inefficiencies.

\subsection{Privacy Issues}
As noted above, parameter-based FL works have been shown to be highly susceptible to privacy leakage~\cite{zhu2019deep, geiping2020inverting}. Distillation-based FL works with recursive model exchanges involved ~\cite{lin2020ensemble, zhu2021data} also post privacy risk.
Utilizing unlabeled public data during distillation has proven to be effective in protecting private local data from attackers~\cite{hamm2016learning}. PATE~\cite{papernot2016semi} also suggests that restricting the student network's access to the teacher's network and training with  non-overlapping public datasets can further guarantee privacy protection. Unlike PATE, which uses topmost local votes to train the central model, we quantize and add noise on logits for aggregation and distillation. This retains more local expertise information and therefore improves the utility of the target model without sacrificing the protection of private data.

\section{Method}

In a federated learning setting with $K$ local nodes, each local note hosts a private, labeled dataset $\mathcal{D}^{k}=\{(\bm x_i^k, y_i^k)|i=1,\ldots,|\mathcal{D}^k|\}$. A shared, unlabeled public dataset $\mathcal{D}^{0}=\{\bm x_i^0|i=1,\ldots,|\mathcal{D}^0|\}$ is accessible by the central server and all local nodes. In the first stage of FedKD, the model at each local node $k$ is initialized with model parameters $\theta^k$ by training with its own local private data $\mathcal{D}^k$. Note that FedKD is agnostic to the type of neural network architecture, and hence each local node can have its own specialized architecture suited for the particular distribution of its local data.

In the second stage, the local, private datasets are first disconnected from local training servers to minimize the risk of any data leakage and to protect privacy. The public dataset $\mathcal{D}^0$ that is hosted on the server and deployed at each local node is then used for one-way knowledge distillation from the local nodes to the server. Local models $\theta^k$, together with the central model $\theta^\mathrm{s}$ on the server, constitutes a student-teacher knowledge transfer configuration.  The teacher here is an ensemble of multiple local models, one at each local node. The following sections introduce our privacy-preserving ensemble and distillation schemes for various tasks.

\subsection{Privacy-Preserving Ensemble}
The private dataset is denoted $\mathcal{D}^k=\{(\bm x_i^k, y_i^k)|i=1, \ldots, |\mathcal{D}_k|\}$ ($k \in \mathcal{K}$), where $ \mathcal{K} = \{1,...,K\}$, $y^k \in \mathcal{C}^k$, $\mathcal{C}^k$ is the set of existing classes in the dataset $\mathcal{D}^k$, and $\mathcal{C}^k \subset \{1,\ldots,C\}$ ($C$ is the number of classes across all local nodes). Let $z_i^{ck}=f(\bm x_i^0,\theta_k, c)$  be the logits of a public data sample $\bm x_i^0$ corresponding to class $c \in \mathcal{C}^k$, produced by the model at local node $k$, where $c \in \{1,\ldots,C\}$. We omit $i$ in the following descriptions for simplicity. The conventional aggregation $\widehat{z}^{c}= \frac{1}{|\mathcal{K}|}\sum_{k \in \mathcal{K}} z^{ck}$ takes an average of all teachers' logits.
However, under the FL setting with a high degree of heterogeneity, a conventional ensemble algorithm is not appropriate primarily due to its inability to cope with the more general scenarios when local nodes are not sharing the exact same set of target classes. To take this into consideration, we introduce an importance weight $\omega$ for each local node to reflect the distribution of the local private data:
\begin{equation}\label{eqweight}
\omega_k^c = \frac{N_{k}^c}{\sum_{k \in \mathcal{K}} {N_{k}^c}},
\end{equation}
where for single-label classification, $N_{k}^c = \sum_{i=1}^{|\mathcal{D}_k|} (y_i^k=c)$ denotes the number of samples of class $c$ used in training the model at local node $k$. 

Inspired by PATE \cite{papernot2016semi}, we perturb the locally computed logits with a quantized and noisy ensemble for a stronger privacy guarantee: 
\begin{equation}\label{eqensemble}
\widehat{z}^{c} = \sum_{k \in \mathcal{K}} {\omega_k^c \cdot Q(z^{ck}; S)} +Lap(\frac{1}{\gamma}), 
\end{equation}
where $Q(\cdot, S)$ is the quantization function with $S$ as quantization scale, and $Lap(\frac{1}{\gamma})$ is the Laplacian distribution with location $0$ and scale $\frac{1}{\gamma}$. $\gamma$ is a privacy parameter to trade off between privacy-preserving capability and accuracy. A smaller $\gamma$ (i.e., higher noise level) results in stronger privacy guarantee and relatively lower accuracy.

To achieve better communication efficiency, we apply uniform quantization to floating point logits so they occupy fewer bits:
\begin{equation}\label{eq:quan_eq}
    Q({z}^{ck}; S) = q_{s}, ~\text{if}~{z}^{ck} \in (q_{s-1}, q_{s}],
\end{equation}

We determine the quantization intervals $(q_{s-1}, q_{s}]$ with
$q_s = \frac{2 (s-1) {z}^\text{max}}{S-1} - {z}^\text{max} (s=1, \ldots, S)$, 
where ${z}^\text{max}=\operatorname{max}_{i,c,k} |{z}_i^{ck}|$ is the maximum absolute value of all the logits across public data samples $i=1,\ldots,|\mathcal{D}_0|$ and classes $c=1,\ldots,C$. Thus, Equation \eqref{eq:quan_eq} becomes:
\begin{equation}
    Q({z}^{ck}; S) = \lceil \frac{S \cdot {z}^{ck}}{2 {z}^\text{max}} \rceil \cdot  \frac{2 {z}^\text{max}}{S},
\end{equation}
where a smaller $S$ sacrifices more logits precision, while maintaining a higher level of privacy.

During ensemble, we protect the private data at each local node by: (1)  transferring only the final prediction inferred with the non-proprietary public data $\mathcal{D}_0$;  and (2) perturbing the local outputs with quantization and random noise .

\begin{algorithm}[tb]
 \begin{algorithmic}
    \STATE {\bfseries Input:} Labeled private datasets $\{\mathcal{D}^k|k \in \mathcal K\}$ ($\mathcal K=\{1,\ldots,K\}$), unlabeled public data $\mathcal{D}^0$, central model $\theta^\mathrm{s}$, local models $\{\theta^k| k \in \mathcal K\}$,  $T$ distillation steps, batchsize $B$, quantization scale $S$, privacy hyperparameter $\gamma$.
    \STATE {\bfseries Local Training:} \\ Train each local model $\theta^k$  with private data $\mathcal{D}^k$.
    \STATE {\bfseries Logits Ensemble:} \\
    \FOR {each sample $\bm{x}_i^0$ in $\mathcal{D}^0$}
        \FOR {each local $k \in \mathcal K$ }
        \STATE $\bm{z}_i^k$ $\leftarrow f(\bm{x}_i^0, \theta^k)$
        \ENDFOR
        \STATE $\widehat{\bm{z}}_i \leftarrow$ aggregate $\{\bm z_i^k; S, \gamma |k \in {\mathcal K} \}$  \quad $\triangleright$ {Eq.~\ref{eqensemble}}
    \ENDFOR
    \STATE {\bfseries Distillation:} 
    \FOR{each distillation step $t=1,...,T$}
    \STATE $\bm{x}^0$ $\leftarrow$ a batch of public data from $\mathcal{D}^0$ with size $B$
    
    \STATE $\widetilde{\bm z} \leftarrow f(\bm{x}^0, \theta^\mathrm{s})$ 
    \STATE Update the central model: $\theta^\mathrm{s}$ $\leftarrow$ ${\theta^\mathrm{s}} - \frac{1}{B}\nabla_{\theta^\mathrm{s}} \Larg$ \quad $\triangleright$ {Eq.~\ref{eqloss1}}
    \ENDFOR
\end{algorithmic}
\caption{Federated Knowledge Distillation (FedKD)}
\label{alg}
\end{algorithm}

\subsection{One-shot Distillation}
Conventional knowledge distillation aggregates all teachers' soft labels subject to the Kullback-Leibler divergence:
\begin{equation}\label{eq1}
    \Larg =  \sum_c  p^c log{\frac{ p^c}{ q^c}},
\end{equation}
where $p^c$ and $q^c$ denote the probabilities of a sample of class $c$ for the teacher and student models, respectively. 
The aggregated logits $\widehat{z}^{c}$ can be viewed as teacher knowledge, and the output logits of the central model $\widetilde {z}^c=f(\bm x_0,\theta_\mathrm{s}, c)$ can be viewed as student knowledge. Without loss of generality, we denote the activation by $p^c = \sigma(\widehat{z}^{c})$ and $q^c = \sigma(\widetilde z^c)$. For single-label classification, we obtain the probabilities using softmax activation:
\begin{equation}
\label{eq2}
p^c= \sigma(\widehat{z}^{c})=\frac{e^{\widehat{z}^{c}/\tau}}{\sum_c{e^{\widehat {z}^{c}/\tau}}}, 
~q^c =\sigma(\widetilde{z}^{c})=\frac{e^{\widetilde {z}^c/\tau}}{\sum_c{e^{\widetilde {z}^c/\tau}}},
\end{equation}
where $\tau$ is a temperature parameter.   Hinton et al. \cite{hinton2015distilling} showed that minimizing Eq.~\ref{eq1} with high $\tau$ is equivalent to minimizing the $\ell_2$ error between the teacher and student logits, thereby relating cross-entropy minimization to matching logits.

Based on the observations above by Hinton et al. \cite{hinton2015distilling}, we consider the case of $\tau \rightarrow \infty$ so the loss can be written as:
\begin{equation} \label{eqloss1}
\Larg = {\| \widetilde{\bm z}- \widehat{\bm z}\|},
\end{equation}
where $\widetilde{\bm z} = [\widetilde{z}^1,... \widetilde{z}^C]$, and $\widehat{\bm z} = [\widehat{z}^1,... \widehat{z}^C]$.

Note that we use one-shot offline distillation where the local nodes predict with each public data sample only once, and the predicted logits are used to train the central model iteratively. This distillation strategy (1) provides a higher privacy guarantee by executing fewer queries to the local model (limiting the access to local knowledge); and (2) eliminates the iterative and repetitive communication requirement of synchronous updates, improving communication efficiency and flexibility. The overall process is described in Algorithm~\ref{alg}.

\begin{table*}[]
\centering
\fontsize{9.0pt}{10.0pt} \selectfont
{
\begin{tabular}{c|ccc|ccc}
\toprule
\multirow{3}{*}{\textbf{Method}} & 
\multicolumn{3}{c|}{CIFAR-10} & 
\multicolumn{3}{c}{CIFAR-100} \\ 
\cmidrule(lr){2-7}
    &\multicolumn{2}{c}{\textbf{Accuracy}(\%) $\uparrow$} &{\textbf{Bandwidth}}  &\multicolumn{2}{c}{\textbf{Accuracy}(\%) $\uparrow$} &{\textbf{Bandwidth}} \\
    &$\alpha=1$ &$\alpha=0.1$ &(GB)$\downarrow$ &$\alpha=1$ &$\alpha=0.1$ &(GB)$\downarrow$ \\
    \midrule \midrule
FedAvg~\cite{mcmahan2017communication}
&
78.57 $\pm$0.22 & 
68.37$\pm$0.50 & 58
&
42.54$\pm$0.51 &
36.72$\pm$1.50 &63\\
FedProx~\cite{li2018federated}
&
76.32 $\pm$1.95 & 
68.65$\pm$0.77 & 58
&
42.94$\pm$1.23 & 
35.74$\pm$1.00 &63\\
FedAvgM~\cite{hsu2019measuring}
&
77.79$\pm$1.22 & 
68.63$\pm$0.79 & 58
&
42.83$\pm$0.36 & 
36.29$\pm$1.98 &63\\
FedDF~\cite{lin2020ensemble}
&
80.69$\pm$0.43 & 
71.36$\pm$1.07 &58 
& 
\bf{47.43}$\pm$0.45 & 
39.33$\pm$0.03 &63\\
FedGEN~\cite{zhu2021data} 
&
80.31$\pm$0.97 & 
68.13$\pm$1.37 &58 
&
45.97$\pm$0.23& 
35.97$\pm$0.31 &63\\
FedMD~\cite{li2019fedmd}
&
80.37$\pm$0.37 & 
\bf{69.23}$\pm$1.31 &6.24
&
45.83$\pm$0.58 & 
38.86$\pm$0.78 &160\\\midrule
\textit{Standalone}
&
{61.11}$\pm$24.90 & 
{28.99}$\pm$27.24 &- 
&
{27.49}$\pm$14.76 & 
{16.31}$\pm$15.75 &-\\
\midrule
FedKD 
&
\bf{80.98}$\pm$0.11 & 
{65.46}$\pm$3.45 &\bf{0.078}
&
{45.55}$\pm$0.38& 
\bf{40.61}$\pm$2.54 &\bf{2} \\
\bottomrule
\end{tabular}}
\caption{Comparisons on the CIFAR-10 and CIFAR-100 datasets with ResNet-8 when $K$=20. Our FedKD uses $S$=200, $\gamma$=1 for knowledge ensemble, while the competing methods use the setting in FedDF~\cite{lin2020ensemble} with 100 rounds and a sampled fraction as 1 at each communication round. \textit{Standalone}: mean/std performance of all local models. Both logits and parameters are of type float64 for bandwidth calculation.}
\label{tab:cifarcompare}
\end{table*}

\subsection{Cross-domain Analysis}
We argue that with cross-domain public data our framework can distill knowledge from multiple locals with generalizability. In this section we present a performance bound for the aggregated central model, which is built upon prior arts from domain adaptation~\cite{ben2010theory}.

Let the input space be $\mathcal{X}$, $\mathcal{D}^S$ and $\mathcal{D}^T$ be source and target domain respectively,
We denote the ground-truth labeling function as $g$ and the hypothesis function as $f$, we get the error as $\epsilon_{\mathcal{D}^S}(h,g)= \mathbb{E}_{x \sim \mathcal{D}^S}[|h(x)-g(x)|]$. We denote the risk of $h$ on $\mathcal{D}^S$ and $\mathcal{D}^T$ as $\epsilon_{\mathcal{D}^S}$ and $\epsilon_{\mathcal{D}^T}$. 
\cite{ben2010theory} introduces $\mathcal{H}$-divergence to evaluate the distance between two domain distributions $\mathcal{U}$, $\mathcal{U'}$ on the a hypothesis space $\mathcal{H}$.
$\mathcal{H}$-divergence is defined as $d_{\mathcal{H}}(\mathcal{U}, \mathcal{U}') = 2\operatorname{sup}_{A \in \mathcal{A}_{\mathcal{H}}}|\operatorname{Pr}_{\mathcal{D}}(A) - \operatorname{Pr}_{\mathcal{D}'}(A)|$, where $\mathcal{A}_{\mathcal{H}}$ denotes a collection of subsets of $\mathcal{X}$ which support the hypothesis in $\mathcal{H}$. 
The symmetric different space is defined as $\mathcal{H} \Delta \mathcal{H}= \{ h(x) \bigoplus h'(x)| h, h' \in \mathcal{H}\}$ ($\bigoplus$ represents the XOR operation). For the generalizability between two domains, we have the following theorem ~\cite{blitzer2007learning}: 

{Theorem 1. Generalization bounds.} \textit{Let $\mathcal{H}$ be a hypothesis space of VC dimension $d$, $\mathcal{U}^S$ and $\mathcal{U}^T$ be unlabeled samples of size $N$ each, drawn from $\mathcal{D}^S$ and $\mathcal{D}^T$ respectively.  For any $h \in \mathcal{H}$ and $\delta \in (0,1)$, the following holds with probability at least $1-\delta$ (over the choice of the samples):}
\begin{equation}
\begin{aligned}
     \epsilon_{\mathcal{D}^T}(h) \leq & \epsilon_{\mathcal{D}^S}(h) + \frac{1}{2} {d}_{\mathcal{H} \Delta \mathcal{H}} (\mathcal{U}^S, \mathcal{U}^T) \\
     &+ 4 \sqrt{\frac{2d\operatorname{log}(2N)+\operatorname{log}(\frac{2}{\delta})}{N}} + \lambda,
\end{aligned}
\end{equation}
\textit{where $\lambda=\epsilon_{\mathcal{D}^S}(h^*)+\epsilon_{\mathcal{D}^T}(h^*)$ and $h^*$ is the ideal joint hypothesis minimizing the combined error: $h^* = \operatorname{argmin}_{h \in \mathcal{H}} \epsilon_{\mathcal{D}^S}(h^*)+\epsilon_{\mathcal{D}^T}(h^*)$.}

In our case, $\mathcal{D}^S$ is the domain of private data distributed across $K$ local nodes: $\mathcal{D}^S = \{ \mathcal{D}^k|k \in \mathcal{K} \}$, and $\mathcal{D}^T$ = $\mathcal{D}^0$ is the domain of public data. We assume $|\mathcal{D}^0|=N$, $\sum_{k \in \mathcal{K}} |\mathcal{D}^k|=N$. 
Given the local model $h_{\mathcal{D}^k}$ trained on data $\mathcal{D}^k$, we learn central model $h_{\mathcal{D}^0}$ from public data ${\mathcal{D}^0}$ through weighted aggregation: $h_{\mathcal{D}^0} = \sum_{k \in \mathcal{K}} \omega_k (h_{\mathcal{D}^k}+n_k(\gamma))$,  where $\sum_{k \in \mathcal{K}} \omega_k=1$, and $n_k(\gamma)$ is the introduced noise parameterized by $\gamma$ to strengthen the privacy. We have the following weighted noisy generalization bound:
\begin{equation}
\begin{aligned}
    \epsilon_{\mathcal{D}^0}(h_{\mathcal{D}^0}) \leq & \epsilon_{\mathcal{D}^S}\left(\sum_{k \in \mathcal{K}}{\omega_k (h_{\mathcal{D}^k}+n_k(\gamma))}\right) + \lambda_\omega\\ 
    &+  \sum_{k \in \mathcal{K}} \omega_k\left( \frac{1}{2} {d}_{\mathcal{H} \Delta \mathcal{H}} (\mathcal{U}^k, \mathcal{U}^0) \right) \\
    &+ 4 \sqrt{\frac{2d\operatorname{log}(2N)+\operatorname{log}(\frac{2}{\delta})}{N}}.
\end{aligned}
\end{equation}

\subsection{Extension to Other Tasks}
While Eq.~\ref{eq2} corresponds to the single-label classification scenario, our method is also extensible to multi-label classification. In this case, the private data notation from above is changed to $\mathcal{D}^k=\{(\bm x_i^k, \bm{y}_i^k)| i=1,\ldots,|\mathcal{D}^k|\}$ with $\bm{y}_i^k \in \{-1,0,1\}^c$ where -1, 0, and 1 indicate unknown, negative, and positive for class $c \in {1,...,C}$, respectively. We have made two other modifications: first, a sigmoid is used as the activation instead of softmax so $p^c = \sigma(\widehat{z}^{c})$ and $q^c = \sigma({\widetilde z}^{c})$; second, in Eq.~\ref{eqweight}, we define $N_{k}^c= \sum_{i=1}^{|\mathcal{D}^k|}(\bm{y}_i^k(c)= 1 )$ as the number of samples labeled as class $c$ for training the model of local node $k$.

\begin{table}
\begin{center}
\fontsize{9.0pt}{10.0pt} \selectfont
{
\begin{tabular}{c|c|c|c|c}
\hline
\textit{aggregation scheme} &baseline &Eq.~\ref{eqweight} &Eq.~\ref{eqweight} &Eq.~\ref{eqweight}\\
\textit{logits distillation} &$\tau$=$\infty$ &$\tau$=3 &$\tau$=$\infty$ &$\tau$=$\infty$ \\
\textit{\# local prediction} &$|\mathcal{D}^0|$ &$|\mathcal{D}^0|$ &$|\mathcal{D}^0|$ &50$\times |\mathcal{D}^0|$\\\hline
Accuracy(\%)$\uparrow$ &79.92 &80.01 &80.98 &\bf{81.89}\\ \hline
Bandwidth (GB) $\downarrow$ &\multicolumn{3}{c|}{0.078} &3.91 \\ \hline
\end{tabular}}
\end{center}
\caption{Ablation study on CIFAR-10 with ResNet-8, $K$=20, $\alpha$=1, $S$=200, $\gamma$=1. With the commonly used distillation scheme (temperature $\tau = 3$) as baseline, we show the comparison on different ensemble and distillation schemes. $|\mathcal{D}^0|$ indicates the number of samples in the public dataset $\mathcal{D}^0$, and 50$\times |\mathcal{D}^0|$ indicates local model predicts 50 times on each sample of $\mathcal{D}^0$ with different augmentation seeds.}
\label{tab:c10ab}
\end{table}

\section{Experiments}
We conduct experiments on natural image classification (single-label), medical image classification (multi-label), and extensive experiments on text classification. We construct local training sets using heterogeneous data splits with a Dirichlet distribution as in prior works \cite{hsu2019measuring}. The value of $\alpha$ controls the degree of non-IID-ness.  An $\alpha$ of positive infinity  indicates identical local data distributions, and a smaller $\alpha$ indicates higher non-IID-ness. 

\begin{figure}
\centering
\subfloat[][Varying $\frac{1}{\gamma}$ and $|\mathcal{D}^0|$.]{\includegraphics[width=0.5\columnwidth]{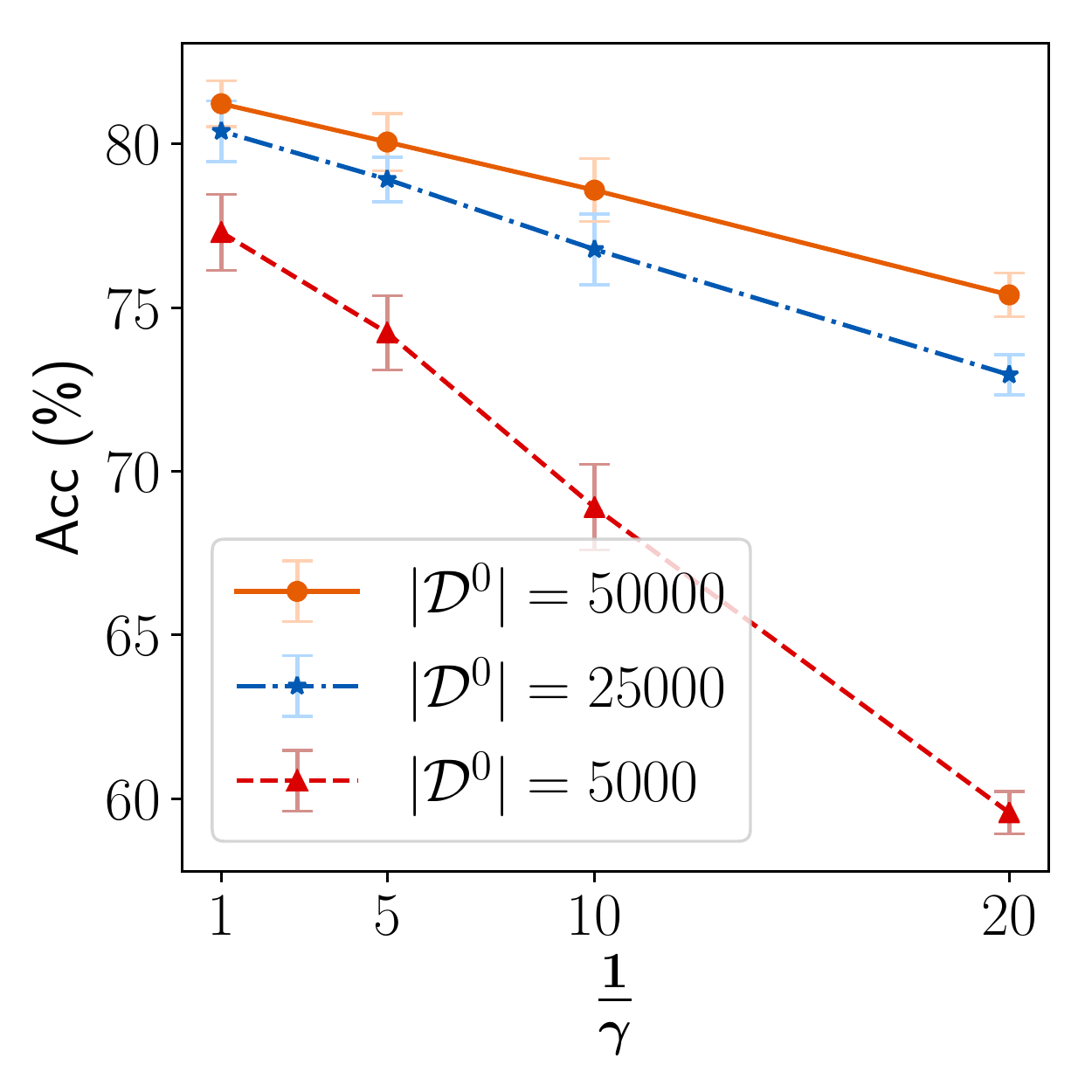}}
\subfloat[][Varying $S$ and $\frac{1}{\gamma}$. ]{\includegraphics[width=0.5\columnwidth]{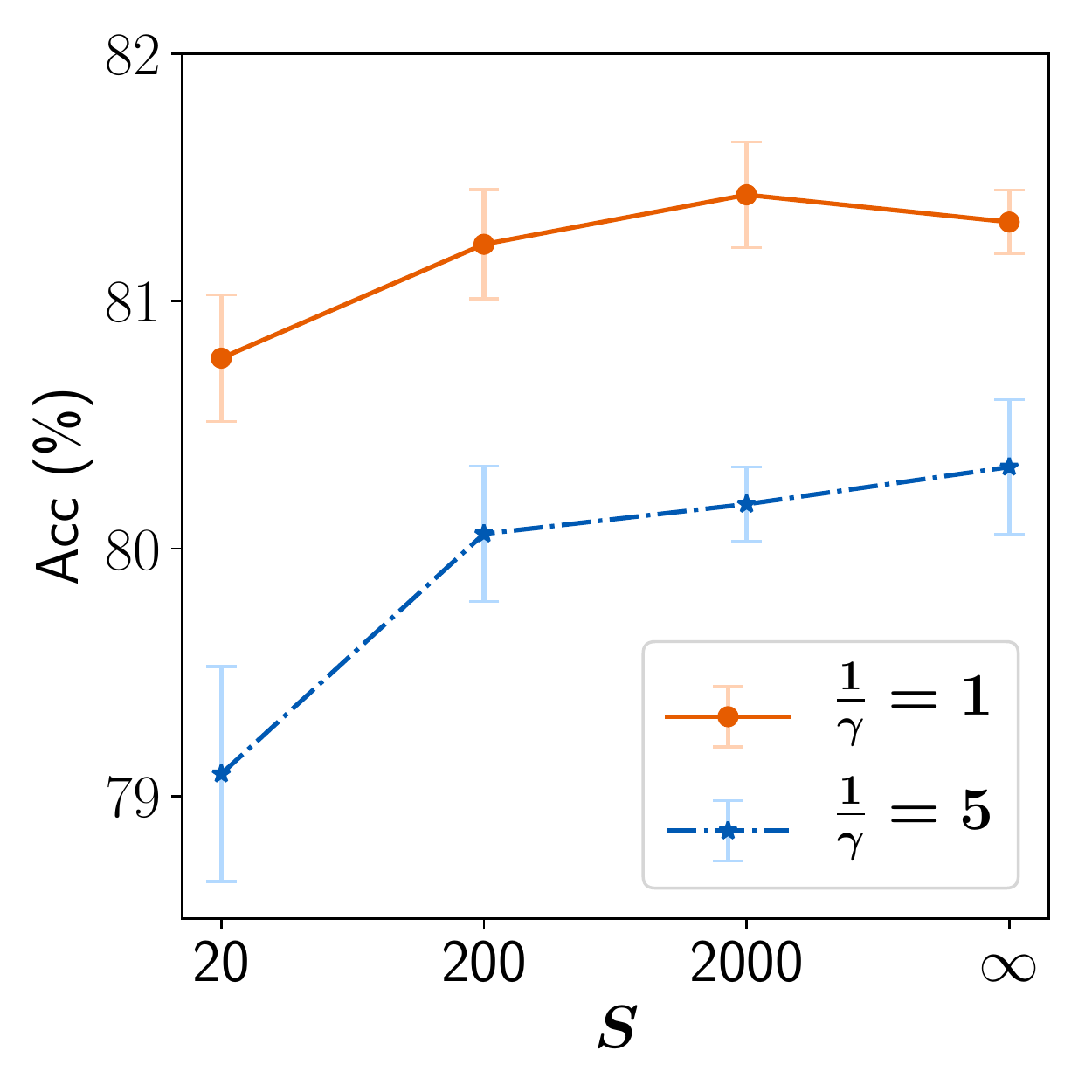}}
\caption{Ablation study on the CIFAR-10 dataset ($K$=20, $\alpha$=1) with varying public data size $|\mathcal{D}^0|$, noise $\frac{1}{\gamma}$, and quantization scale $S$. }
\label{fig:c10ab}
\end{figure}

\begin{table*}
\centering
\fontsize{9.0pt}{10.0pt} \selectfont
{
\begin{tabular}
{c|cc|cc}
\toprule
Method &Private data $\mathcal{D}^k$  &Test &$K=3$ &$K=5$ \\  \hline \hline
\multirow{2}{*}{FedKD (Single-domain)} &CXR14 &CXR14   &75.02 &74.80  \\ 
&Xpert &Xpert  &82.41  &82.35  \\ \hline \hline
\multicolumn{3}{c|}{} &$K=2$ &$K=6$    \\\hline
\multirow{2}{*}{FedKD (Cross-domain)} &CXR14+Xpert &CXR14 &79.03 &76.13  \\
&CXR14+Xpert &Xpert &79.77 &80.91 \\\bottomrule
\end{tabular}}
\caption{Multi-label classification experiments on chest-x-ray images with single/cross domain private data. We report the test mAUC (\%) on NIH CXR14 and CheXpert over 12 and 8 classes respectively.}
\label{tab:xray}
\end{table*}


\subsection{CIFAR10/100 Classification}
For natural image classification task we use CIFAR-10/100~\cite{krizhevsky2009learning} as datasets. To keep consistency with the prior arts, we use the same experimental settings as in FedDF~\cite{lin2020ensemble}: CIFAR-100 as unlabeled public data when CIFAR-10 as private data, and downsampled version of ImageNet ($32 \times 32$) \cite{deng2009imagenet} as unlabeled public data corresponding to CIFAR-100 as private data. For each experiment, we sample over three different random seeds as private data split for local training. 
We report the average accuracy metrics on CIFAR-10 and CIFAR-100 test set corresponding to its private data respectively.  

\subsubsection{Implementation Details.}
Following \cite{lin2020ensemble,gong2021ensemble}, we use ResNet-8 as backbone. We train each local model individually with SGD and Cosine Annealing \cite{loshchilov2016sgdr}, decreasing the learning rate from 0.0025 to 0.001 in 500 epochs with a batch size of 16. For distillation, we use the Adam optimizer,  a constant learning rate of 1e-3, and a batch size of 512.  We use 200 and 10 epochs for CIFAR-10  and  CIFAR-100 respectively. The weight decay is 3e-4 and 0 for local training and distillation, respectively.

\subsubsection{Results.}
The comparison in Table~\ref{tab:cifarcompare} shows that our method achieves a significantly stronger privacy guarantee as well as a far better communication efficiency compared to prior arts, without sacrificing accuracy. On CIFAR-10 ($\alpha=1$) and CIFAR-100 ($\alpha=0.1$), our method demonstrates better accuracy with significantly lower communication cost than the prior arts. On CIFAR-10 ($\alpha=0.1$) and CIFAR-100 ($\alpha=1$), our method achieves the best performance-bandwidth trade-off compared with the prior arts. More importantly, our method does not share any locally trained model parameters and further adds noise perturbation on the transferred product of non-sensitive public data, demonstrating stronger privacy guarantee than the prior arts. 

\subsubsection{Ablation Studies.}
We perform ablation studies to validate the efficacy of our ensemble and distillation strategy and show the results in Table~\ref{tab:c10ab}. The extensive experiments in Table~\ref{tab:c10ab} show the distillation accuracy can be improved by a large margin with more access to local information (e.g., local models predicted on dynamically augmented public data multiple times). For an accuracy-privacy trade-off, we restrict that local model to only predict each public sample once in our method. Besides, we do ablation study with different temperatures $\tau$ for logits distillation ~\cite{hinton2015distilling}.

In Figure~\ref{fig:c10ab} we study the impact of quantization/noise on the accuracy for different sized public datasets. The left figure suggests that increased noise degrades the ensemble distillation performance,  but a (unlabeled) larger public dataset can substantially improves the robustness to noise perturbation. 
We observe from the right figure that the distillation results are insensitive to data precision, which is also observed in prior work \cite{shazeer2017outrageously}. Thus we use $S=200$ and $\gamma=1$ as default setting in the following experiments.

\begin{table*}
\centering
\fontsize{9.0pt}{10.0pt} \selectfont
{
\begin{tabular}{c|ccccc|c|ccccc|c}
\toprule
&\multicolumn{6}{c|}{Homogeneous} &\multicolumn{6}{c}{Heterogeneous} \\\cmidrule(lr){2-13}
&CM &ED &CS&AT &PE &{\textbf{mAUC}} &CM &ED &CS&AT &PE &{\textbf{mAUC}} \\\midrule
\textit{Standalone} &
\renewcommand\arraystretch{0.8} \begin{tabular}{@{}c@{}} 78.57 \\ \footnotesize{$\pm$2.27}\end{tabular} &
\renewcommand\arraystretch{0.8} \begin{tabular}{@{}c@{}} 85.82 \\ \footnotesize{$\pm$1.95}\end{tabular} &
\renewcommand\arraystretch{0.8} \begin{tabular}{@{}c@{}} 88.16 \\ \footnotesize{$\pm$2.12}\end{tabular} &
\renewcommand\arraystretch{0.8} \begin{tabular}{@{}c@{}} 79.87 \\ \footnotesize{$\pm$4.22}\end{tabular} &
\renewcommand\arraystretch{0.8} \begin{tabular}{@{}c@{}} 84.60 \\ \footnotesize{$\pm$1.58}\end{tabular} &
\renewcommand\arraystretch{0.8} \begin{tabular}{@{}c@{}}  83.67 \\ \footnotesize{$\pm$1.24}\end{tabular} &
\renewcommand\arraystretch{0.8} \begin{tabular}{@{}c@{}} 69.12 \\ \footnotesize{$\pm$5.15}\end{tabular} &
\renewcommand\arraystretch{0.8} \begin{tabular}{@{}c@{}} 82.63 \\ \footnotesize{$\pm$3.48}\end{tabular} &
\renewcommand\arraystretch{0.8} \begin{tabular}{@{}c@{}} 83.26 \\ \footnotesize{$\pm$2.74}\end{tabular} &
\renewcommand\arraystretch{0.8} \begin{tabular}{@{}c@{}} 70.71 \\ \footnotesize{$\pm$0.63}\end{tabular} &
\renewcommand\arraystretch{0.8} \begin{tabular}{@{}c@{}} 80.32 \\ \footnotesize{$\pm$3.49}\end{tabular} &
\renewcommand\arraystretch{0.8} \begin{tabular}{@{}c@{}} 77.21 \\ \footnotesize{$\pm$1.29}\end{tabular}\\
\textit{Public-only} &67.34 &79.76 &79.24 &76.38 &80.37 &82.43 &45.28 &78.03 &77.36 &66.98 &75.43 &68.60 \\
\textit{Centralized} &\textbf{82.88} &\textbf{87.04} &\textbf{91.53} &80.90 &\textbf{87.02} &\textbf{85.88}&75.38 &82.28 &86.37 &\textbf{75.36} &\textbf{85.93} &\textbf{81.07} \\ \cmidrule(lr){1-13}
FedKD &81.81 &86.12 &91.15 &\textbf{83.34} &86.59 &85.81  &\textbf{75.62} &\textbf{82.83} &\textbf{87.95} &74.61 &83.48 &80.90 \\
\bottomrule
\end{tabular}}
\caption{Comparisons of AUCs (\%) on the homogeneous/heterogeneous positive data distribution with $K=5$ and labeled public data. \textit{Standalone}: averaged AUCs of all local models. \textit{Public-only}: training with only labeled public data. \textit{Centralized}: central training with all public and private data. \emph{CM}: Cardiomegaly, \emph{ED}: Edema, \emph{CS}: Consolidation, \emph{AT}: Atelectasis, \emph{PE}: Pleural Effusion. }
\label{tab:xpert1}
\end{table*}

\begin{table*}
\centering 
\fontsize{9.0pt}{10.0pt} \selectfont
{
\begin{tabular}
{cc|ccc|cc}
\toprule
& &FedAvg &FedDF &FedKD &\textit{Standalone} &\textit{Centralized} \\\midrule
\multirow{2}{*}{AG News} &Accuracy (\%) $\uparrow$  &91.98 &92.57 &\bf{92.58} &86.30$\pm$5.21 &93.11\\
&Bandwidth(MB) $\downarrow$ &10217 &10235 &\bf{36.6} &- &-\\ \midrule
\multirow{2}{*}{SST2} &Accuracy (\%) $\uparrow$ &87.13 &{88.51} &\bf{91.50} &74.80$\pm$5.05 &90.07 \\ 
&Bandwidth(MB) $\downarrow$ &10217 &10221 &\bf{10.3} &- &-\\\midrule
\multicolumn{2}{c|}{Privacy (NO shared Param.)} &\xmark &\xmark &\cmark &- &-\\
\bottomrule 
\end{tabular}}
\caption{Comparisons on AG News and SST2 datasets with $K$=10 under the same experiment setting. \textit{Standalone}: mean $\pm$ std of local models trained with individual private data. \textit{Centralized}: centralized training all local private data.}
\label{nlp}
\end{table*}

\subsection {Chest X-Ray Image Classification}
Although mainstream FL methods  experiment exclusively with private data from the same dataset (domain), this is typically not realistic in practical applications.  For example, data acquired at different hospitals may come from different sources. We thus consider a more general heterogeneous setting where the private data at different local nodes and the unlabeled public data all come from different domains. 

Here we implement multi-label classification on chest-x-ray images, using the NIH CXR14~\cite{wang2017chestx} and CheXpert~\cite{irvin2019chexpert} datasets to represent different domains for private data. 
We ignore ambiguous categories (Effusion, Pleural Effusion, Pleural Other and Support Device), remaining a total of 14 annotation classes, of which NIH CXR14 has annotations for 12 classes and CheXpert for 8 classes, with 6 overlapping classes. So there are totally 86,524 images come from NIH CXR14 and 64,346 images come from CheXpert dataset. For each dataset, we randomly sample 90\% for training and the rest 10\% for validation. 
We use 26,684 images from the RSNA Pneumonia Detection Challenge~\cite{rsnadata} without using their labels as public data. 

\subsubsection{Implementation Details.}
We use ResNet-34 as the backbone.
For local training, we use a batch size of 32, same data augmentation strategies as in prior work \cite{ye2020weakly}. We train each local model individually with SGD and Cosine Annealing, decreasing the learning rate from 1e-4 to 1e-6 in 50 epochs. For distillation, we use SGD and a constant learning rate of 1e-3 and 50 epochs. 
For samples with multiple classes labeled as positive, we choose the most infrequent one (the class with least positive samples) as its label for the Dirichlet data split.
In the setting with cross-domain private data (two datasets as private data), each dataset is distributed to $K_d=K/2$ local nodes when there is a total of $K$ local nodes.  
\subsubsection{Results with Unlabeled Public Data.}
In Table~\ref{tab:xray}, we first study the hyper-parameters $K$ with $\alpha=1$, $S=200$, $\gamma=1$ and local data from a single dataset (domain). It shows larger numbers of locals $K$ negatively affects the distillation performance. Table~\ref{tab:xray} also shows cross-domain, cross-site evaluations using both datasets as private data, with a total of $K$ local nodes ($K_d=K/2$ for each dataset, and each node hosts data from only one of the datasets). 
We can see that the introduction of additional cross-domain local nodes will help to improve the performance of the source domain: CXR14 ($K=3$) as private datasets achieves 75.02\% on CXR14 test set while CXR14+Xpert ($K=6$) as private datasets achieves 76.13\%. 
Note that the model trained with this cross-domain setting is capable of classifying all 14 classes, whereas training with a single domain can only classify 12 and 8 classes, respectively.

\subsubsection{Ablation Studies on Heterogeneity with Labeled Public Data.}
In this experiment, we use labeled public data $\mathcal{D}^0=\{(\bm x_i^0, \bm{y}_i^0)| i=1,\ldots,N_0\}$ which is accessible by all local nodes and included in local training along with local private data. Since medical image datasets are usually characterized by a high degree of imbalance (e.g., far more negative samples than positive samples with abnormalities), we study the heterogeneity of the positive distribution, with each local node having an equal number of private samples. We set the number of local nodes to $K=5$ and the data size to $N_k=6000$, $N_0=1000$ and use the official validation set for testing. 
Table~\ref{tab:xpert1} shows results with homogeneous and heterogeneous distributions (w.r.t. positive samples). 
Notably, under both homogeneous and heterogeneous settings, our method achieves results comparable to centralized training on all public and local data.  This can be viewed as an upper bound.

\subsection{Text Classification Tasks}
We evaluate our framework on two text classification datasets: AG News~\cite{zhang2015character} and SST2~\cite{socher2013recursive}.  Following FedDF~\cite{lin2020ensemble}, we use pre-trained DistilBERT~\cite{sanh2019distilbert} as the transformer language model. Local training and distillation takes 100 and 20 epochs, respectively, and the training strategy is the same as FedDF. From Table~\ref{nlp}, we can note that our method gives the best performance on both datasets. On bot AG News and SST2 dataset, our proposed framework achieves superior accuracy and substantially lower communication bandwidth compared to the prior arts. 
More importantly, our method does not share parameters/gradients of local models during communication, which offers much stronger privacy guarantee compared to the prior arts.

\section{Conclusions}
In this work, we propose a novel distillation-based federated learning framework, namely FedKD, which can preserve local data privacy by learning with only unlabeled and domain robust public data.    
To comprehensively address the communication bottleneck, we employ a one-shot and one-way (offline) knowledge distillation process with an efficient ensemble scheme. Experiments on both image classification and text classification tasks demonstrate the efficacy of FedKD with better privacy guarantee compared to prior arts. Given the increasing importance of privacy, we  believe  our  proposed FL method  will  be  a  practical solution  to  facilitate  privacy-preserving decentralized learning across multiple sites in real-world scenarios, especially for medical applications where leveraging valuable local data at different hospitals without exposing proprietary data to privacy risks is essential.  


\section{Acknowledgements}
We thank the reviewers for their
constructive comments and thank Liangchen Song and Barry M. Yao for the discussion and assistance.
\bibliography{egbib}

\end{document}